\documentclass[iop,revtex4]{emulateapj}
\usepackage{lineno}

\newcommand{\kms}{km~s$^{-1}$}
\newcommand{\msun}{${\cal M}_\odot$}

\begin{document}

\renewcommand{\topfraction}{1.0}
\renewcommand{\bottomfraction}{1.0}
\renewcommand{\textfraction}{0.0}

\title{Orbits and structure of  quadruple systems GJ 225.1 and FIN 332}

\author{A. Tokovinin}
\affiliation{Cerro Tololo Inter-American Observatory | NSF's NOIRlab, Casilla 603, La Serena, Chile}
\email{atokovinin@ctio.noao.edu}

\begin{abstract}
Only a handful of quadruple systems with two accurate inner visual orbits are
 known. Architecture of two such systems is studied here to
determine   period  ratios,  mutual   orbit  orientation,   and  other
parameters;   updated   orbital   elements   and  their   errors   are
derived. Gliese 225.1 (HIP 28442) is composed of three K-type and one
M-type dwarfs and has inner orbital periods of 67.2$\pm$0.2 and
23.4$\pm$0.5 yr. Its inner orbits have small mutual inclination and
are likely coplanar with the outer orbit of $\sim$2 kyr period. 
The quadruple system FIN 332 (HIP 92037) consists of four early A type
stars with similar masses and magnitudes. Both its inner orbits with
periods of 27.6$\pm$0.2 and 39.8$\pm$0.4 yr have large eccentricities
(0.82 and 0.84). Their orientation in the sky is remarkably similar.  In
contrast, the outer orbit with a period of $\sim$5 kyr has a large
relative inclination to the inner orbits. Dynamics and formation of these 
quadruple systems are briefly discussed. 
\vspace*{0.5cm} \\
{\bf Accepted for publication in Astronomy Letters. }
\end{abstract}

\maketitle

\section{Introduction}
\label{sec:intro}

Architecture  of hierarchical  stellar systems  bears traces  of their
formation  mechanisms,  still not  fully  understood.  Systems of  2+2
hierarchy considered here (two close pairs on a wide orbit around each
other)  are  rather  typical.  Multiplicity statistics  in  the  solar
neighborhood  shows that  the  incidence of  inner  subsystems in  two
components  of a wide  binary is  correlated rather  than independent,
hinting    that   stars    in   2+2    quadruples    formed   together
\citep{FG67}. Furthermore, known  2+2 quadruples show some correlation
between  their inner  periods and  often have  all four  components of
similar masses  \citep{Tok2008}. However, it is not  clear whether 2+2
quadruples formed preferentially outside-in by successive fragmentation
of  gas at  large,  then small  scales,  inside-out (inner  subsystems
formed first and later became bound  later), or in a common event like
cloud collision \citep{Whitworth2001}.

Recently, \citet{Zasche2019}  studied a  large sample of  2+2 quadruple
systems where  both inner  pairs are eclipsing  (doubly eclipsing). In
these  systems,   both  close   inner  binaries  have   large  
inclinations,  suggesting  (but  not  proving)  coplanarity  of  their
orbits. Even  more surprisingly,  the ratio of  the inner  periods was
found to have preferential values around 1 and 1.5 and to avoid values
around  2, implying  some kind  of a  resonance between  inner orbits.
Considering short  inner periods and  the likely large  outer periods,
existence of dynamical interactions  between inner orbits leading to a
resonance appeared challenging. \citet{Tremaine2020} has picked up the
challenge  and   determined  conditions  where   such  resonances  can
occur. Significant migration of  inner binaries to shorter periods and
their  moderate  separation  seem  to  be  necessary  to  explain  the
resonances found by Zasche et al.

These discoveries  prompted me to  look at wider 2+2  quadruples where
both inner subsystems have known  visual orbits. Apart from the period
ratio, mutual orientation of inner orbits and other parameters contain
information that can throw some  light on the formation mechanisms.  A
prototype of such quadruples is $\epsilon$~Lyr containing four similar
A-type stars. However, the long  periods of its inner subsystems (1800
and 724 yr) do not allow  calculation of accurate orbits, owing to the
lack of sufficient coverage. Situations where both inner visual orbits
in a 2+2  quadruple can be accurately constrained  by observations are
rare; only about a dozen of  such cases are known.  Two 2+2 quadruples
with  accurate  inner orbits  are  featured  here,  GJ 225.1  and  FIN
332.  Their known  orbits are  re-accessed and  updated
using  recent  observations,  and  the  properties of  the  stars  are
determined taking advantage of Gaia parallaxes \citep{Gaia}.



\begin{table*}
\center
\caption{Orbital elements}
\label{tab:orb}
\begin{tabular}{l rrr rrrr rr} 
\hline
System & $P$  & $T$ & $e$ & $a$ & $\Omega$ & $\omega$ & $i$ & $\Sigma M$ & $K_1+K_2$  \\ 
     & yr     & yr  &     & $''$ & $^\circ$ &  $^\circ$ & $^\circ$ & $M_\odot$ & km~s$^{-1}$  \\
\hline
GJ 225.1 A,B  &  67.22 & 1998.08 & 0.462         & 0.953        & 126.2 & 282.9 & 101.9 & 1.14 & 6.1  \\
             &$\pm$0.19 & $\pm$0.21 & $\pm$0.018 & $\pm$0.017 & $\pm$0.2 &$\pm$0.4 & $\pm$0.3 & $\pm$0.01 & \ldots \\ 
GJ 225.1 C,E  &  23.38 & 2015.44 & 0.216         & 0.433        & 146.8 & 178.6 & 98.4 & 0.90 & 10.9  \\
             &$\pm$0.54 & $\pm$0.13 & $\pm$0.013 & $\pm$0.007 & $\pm$0.2 &$\pm$2.2 & $\pm$0.2 & $\pm$0.01 & \ldots \\ 
GJ 225.1 AB,CE  &  2100 & 1934.0 & 0.200         & 11.40        & 147.6 & 85.1     & 100.2 & 2.06 & 3.0  \\
FIN 332 Aa,Ab   & 27.62  & 1994.00  & 0.820      & 0.0911       & 136.0 & 4.6      & 107.9  & 4.66  & 5.2  \\
             &$\pm$0.16 & $\pm$0.23 & $\pm$0.012 & $\pm$0.0009 & $\pm$1.2 &$\pm$4.1 & $\pm$1.2 & $\pm$0.14 & \ldots \\ 
FIN 332 Ba,Bb   & 39.76  & 2005.09  & 0.843      & 0.120       & 119.3 & 305.9     & 106.9 & 5.1  & 3.9  \\
             &$\pm$0.37 & $\pm$0.33 & $\pm$0.020 & $\pm$0.008 & $\pm$1.5 &$\pm$4.1 & $\pm$1.6 & $\pm$1.0 & \ldots \\
STF 2375 A,B & 5000     & 557       & 0.5        & 3.64       & 167.5    &  143.4  & 64.5 &    8.9 & \ldots \\ 
\hline
\end{tabular}
\end{table*}

\section{Gliese 225.1}
\label{sec:GJ225}

A   classical  visual  triple   system  HJ~3823   AB  and   AB,C  (WDS
J06003$-$3102,  HIP 28442,  HD  40887,  GJ 225.1)  turned  into a  2+2
quadruple when a faint satellite near star C was discovered in 2004 by
\citet{Tok2005} using  adaptive-optics imaging in  the infra-red (IR).
The 23-yr  period of this new  subsystem C,E was  initially found from
astrometric perturbations  (wobble) in the motion of  the outer binary
AB,C. Now, 15 years later, the orbit of C,E is well covered by speckle
interferometry at the Southern Astrophysical Research (SOAR) telescope
\citep[][and   references  therein]{SOAR},  allowing   calculation  of
accurate orbital elements.  The orbit of A,B, already well established
from  historic micrometer  measurements, also  benefits  from accurate
modern speckle astrometry.

The Gaia data release 2 \citep{Gaia} provides astrometry of A and C as
though they  were single  stars. The separation  of A,B in  2015.5 was
0\farcs58, and  the Gaia astrometry  of this unresolved pair  has large
errors,  e.g.   parallax 53.97$\pm$0.42  mas.   The  star  C was  also
unresolved by Gaia, but the  magnitude difference of C,E is large, the motion
in  2015.5 was  slow (it  was near  maximum elongation),  hence  the Gaia
astrometry of  C is more  reliable. The parallax of  C, 54.82$\pm$0.08
mas, is  adopted as  the distance to  the system  (18.24\,pc, distance
modulus 1.30 mag).
\subsection{Orbits of GJ 225.1}
\label{sec:orb1}

Table~\ref{tab:orb} gives  updated orbits of the  two inner subsystems
A,B and C,E  of GJ 225.1 and the tentative outer  orbit of AB,CE.  The
elements  and their  errors are  determined by  weighted least-squares
fit,  where  weights are  inversely  proportional  to  the squares  of
adopted measurement errors \citep{orbit}.  The errors are confirmed by
fitting  many artificially  perturbed data  sets. This  procedure also
gives the relative error of the quantity $a^3/P^2$ that determines the
mass sum, accounting for the correlation between $a$ and $P$.

\begin{figure*}
\plotone{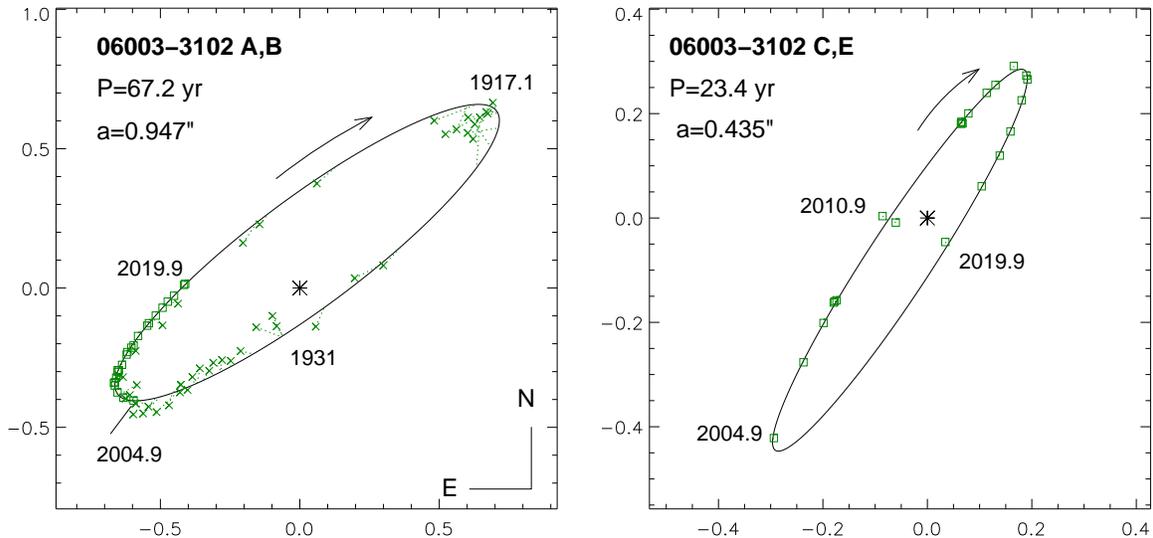}
\caption{Orbits of the inner  subsystems of GJ 225.1. In this and
  following Figures, the primary component of a pair is placed at
 the coordinate origin. The ellipse shows the orbit, with scale in
 arcseconds. Accurate speckle measurements are plotted by squares and
 connected to the respective positions on the orbit by dotted
 line. Less accurate data (mostly historic micrometer measurements) are
 plotted by crosses. 
\label{fig:orb1}
}
\end{figure*}

\begin{figure}
\plotone{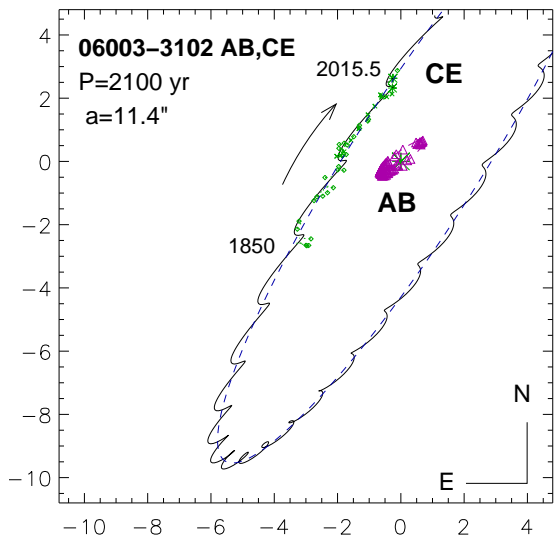}
\caption{Tentative orbit of GJ 225.1 AB,CE. The wobble caused by the
  C,E subsystem is subtracted. The wavy line is the motion of C relative
  to A, subject to the wobble caused by the inner orbit of A,B (plotted on
  the same scale at the center). The dashed line is the center-of-mass
  motion without wobble.
\label{fig:orb1a}
}
\end{figure}

The   inner   orbits  are   plotted   in  Figures~\ref{fig:orb1}   and
\ref{fig:orb1a}.  All orbits  are retrograde (with clockwise rotation)
and   have  a   similar  orientation.    The  last   two   columns  of
Table~\ref{tab:orb} give  the mass sum  computed with the  parallax of
54.82\,mas and  the full  radial velocity (RV)  amplitude $K_1  + K_2$
derived from the orbital elements and the masses estimated below.  The
$a^3/P^2$ ratio is measured with the relative error of 0.056 and 0.007
for A,B  and C,E,  respectively; the full  visual coverage of  the A,B
orbit gives less information on the mass than the still incomplete but
more  accurate  coverage of  C,E.   The  error  of the  Gaia  parallax
contributes a relative error of the mass sum of 0.0044.  The mass sums
are  1.16$\pm$0.06 \msun for  A,B and  0.900$\pm$0.008 \msun  for C,E.
The resulting mass sum of the whole system is 2.06 \msun.

The outer pair  AB,CE has turned by 129.2$^\circ$  from 1850 to 2015.5
(Figure~\ref{fig:orb1a});  the  last  position  is provided  by  Gaia.
Accurate relative positions are measured only after 2004, the rest are
less  accurate micrometer  measurements obtained  from  the Washington
Double  Star  (WDS)  database   \citep{WDS}.   The  motion  is  almost
rectilinear, and the  outer orbit is not constrained  well enough. The
391-yr orbit of  AB,C computed by \citet{Baize1980} is  not viable.  I
fixed  the eccentricity, fitted  all elements,  and refined  the orbit
further by fixing the period  and semimajor axis to values that assure
the correct  outer mass sum of  2.06 \msun.  The outer  orbit given in
Table~\ref{tab:orb}  is  therefore  a  subjective  choice  among  many
potential  orbits that  match the  short  observed arc,  and for  this
reason  no errors are  provided.  This  orbit is  needed mostly  as a
reference for the measurement of the inner mass ratios.

Some measurements  of the outer  pair refer to  A,C, and some  to AB,C
(i.e.   to  the  photo-center  of  the  unresolved  pair  A,B).  These
positions are affected by orbital  motions in both subsystems, and the
resulting wobble  contains information on the inner  mass ratios.  The
wobble  amplitude is proportional  to the  inner orbit  semimajor axis
with  a  scaling  factor  $f  =  q/(1+q)$  in  the  case  of  resolved
measurements of  A,C. The photo-center wobble has  a smaller amplitude
with the  scaling factor $f_\alpha  = f -  r/(1+r)$, where $r$  is the
light ratio in the inner pair.

In an effort to measure the mass ratios, I subtracted the small wobble
caused by  the subsystem C,E  from the outer positions  and determined
the wobble amplitude produced only by the subsystem A,B (the wavy line
in Fig.~\ref{fig:orb1a}). The result is $f_{\rm A,B} = 0.47 \pm 0.02$,
corresponding  to $q_{\rm  A,B}  = 0.89$.   This  mass ratio  slightly
disagrees with the relative  photometry of A,B, which suggests $q_{\rm
  A,B} = 0.84$, the value adopted here.  The resulting wobble
factor $f_{\rm  A,B} =  q_{\rm A,B}/(1+q_{\rm A,B})  = 0.45$  is still
compatible with the measured one.

The  procedure  was repeated  by  subtracting  the  wobble of  A,B  to
determine the mass  ratio in C,E. Given the faintness  of E, $f \approx
f_\alpha$ for this  subsystem.  The result is $f_{\rm  C,E} = 0.24 \pm
0.04$,  hence $q_{\rm C,E}  = 0.32$.   Using the  mass ratios  and the
measured  mass  sums,  masses  of  all four  components  are  computed
(Table~\ref{tab:par1}).   The  relative   motion  between  AB  and  CE
measured by  Gaia was compared to  the motion expected  from all three
orbits.  The agreement  is not as good as  might be expected, probably
because the Gaia astrometry of A is seriously biased by the unresolved
subsystem.   Regretfully,  the  Gaia   astrometry  does  not  help  to
constrain the outer orbit.

\citet{Tok2015} measured in 2008.86 the RVs of unresolved components
AB and CE, both equal to 106.5 \kms. However, the RV amplitudes in
both inner orbits are much larger than the measurement error of
$\sim$0.5 \kms, so this result does not refer to the relative RV of
the centers-of-mass. The spectra of AB have a slightly asymmetric line
profile, suggesting that the RV of the brighter component A was larger
than the RV of B (the predicted RV difference was 5.7 \kms). This
means that the orbital element $\omega_{\rm A,B}$ in
Table~\ref{tab:orb} refers to the ascending node of A. Unfortunately,
the true ascending node of C,E remains unknown. It could be easily
established by RV monitoring of C for several years because the RV
amplitude of C is 2.5 \kms. Spectroscopic orbit of C,E would also
provide accurate measurement of the mass of E.

As the true ascending node of the C,E orbit is not identified, the mutual
inclination $\Phi$ between  orbits of A,B and C,E (i.e. the angle
between the vectors of orbital angular momenta) can take two values,
$20.6^\circ$ or $149.8^\circ$. Small inclination corresponding to
co-rotation appears more likely. Inclinations of the inner orbits to
the uncertain outer orbit are $20.4^\circ$ and  $2.0^\circ$ for A,B
and C,E, respectively; alternative inclinations are $151.2^\circ$ and
$149.8^\circ$. It seems that all three orbits are oriented approximately in one
plane. Mutual inclinations exceeding $39^\circ$ produce Kozai-Lidov
cycles that modulate both inclination and inner eccentricity
\citep{Naoz2016}.    Moderate inner eccentricities support the
near-coplanarity between outer and inner orbits in this system.

Although at present the separation  between AB and CE is comparable to
the  size   of  the  inner  orbits,  hinting   on  possible  dynamical
interaction or  even an instability,  the observed relative  motion of
the wide pair suggests minimum separation  at periastron of $a (1 - e)
=  9\arcsec$,  well  above  the instability  limit  of  $\sim$3\arcsec
according  to the stability  criterion by  \citet{Mardling2001}.  This
hierarchical system  is apparently  not young and  dynamically stable.
The ratio of the two inner periods is 2.87$\pm$0.07.

\begin{figure}
\plotone{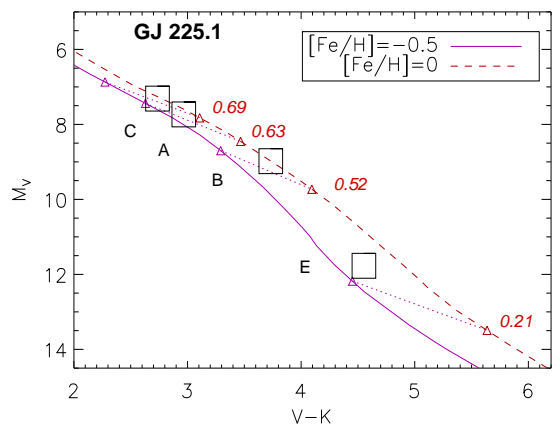}
\caption{Location  of the components of GJ 225.1 on the  color-magnitude diagram.
  The lines  are PARSEC  isochrones \citep{PARSEC} for  1 Gyr  and two
  metallicity values.  Small triangles  connected by dotted lines mark
  measured masses of the components (numbers in italics) on both isochrones. The components
  C, A, B, E (from top down) are plotted by squares.
\label{fig:cmd}
}
\end{figure}

\begin{table}
\center
\caption{Components of GJ 225.1}
\label{tab:par1}
\begin{tabular}{l c c cc  }
\hline
Parameter  & A    & B   & C   & E \\
\hline
$V$ (mag) & 9.04 & 10.29  & 8.62 & 13.08 \\
$V-K$ (mag) & 2.97 & 3.73 & 2.73 & 4.55 \\
${\cal M}$ (${\cal M}_\odot$) & 0.63 & 0.52 & 0.69 & 0.21 \\
\hline
\end{tabular}
\end{table}

\subsection{Photometry and masses of GJ 225.1}
\label{sec:model}

The magnitudes  of individual components in  the IR bands  from $J$ to
$L$ are  measured by \citet{Tok2005}  from the resolved  images.  Gaia
measured the  combined $V$ magnitudes  of AB and  CE as 8.74  and 8.60
mag, respectively (CE is  slightly brighter).  The relative photometry
at SOAR  in the $y$ band  gives the magnitude differences  of 1.25 and
4.46 mag with  the rms scatter of  0.16 and 0.07 mag for  A,B and C,E,
respectively.  Assuming $\Delta  y = \Delta V$, the  $V$ magnitudes of
the  four  stars are  calculated  and  listed in  Table~\ref{tab:par1}
together  with their $V-K$  colors.  Star  C is  the most  massive and
the brightest of all four.

Figure~\ref{fig:cmd}  compares  location  of  the  components  on  the
color-magnitude    diagram   (CMD)    with    the   isochrones    from
\citet{PARSEC}. These dwarf stars  are not evolved.  The components C,
A, and  B are consistent with  normal dwarfs of measured  masses and a
slightly sub-solar metallicity,  [Fe/H]$\approx$$-$0.25 dex. The least
massive star E appears somewhat  brighter and bluer than expected. The
discrepancy  is probably  explained by  imperfect isochrones  for such
low-mass stars. The anomalously blue  $J-K$ color index of E was noted
by \citet{Tok2005}.

The fast  proper motion (PM)  and the large  RV mean that  this system
belongs  to  the  thick  disk Galactic  population.   Considering  all
orbits,  the PM  of the  system's center-of-mass  should  be $(-461.8,
+415.9)$ mas/yr.  Together with the parallax and RV, this leads to the
heliocentric  velocity  of $(U,V,W)  =  (-86.5,  -47.2, -67.3)$  \kms.

\section{Finsen 332}
\label{sec:FIN332}

The second  resolved visual quadruple system considered  here is known
as  WDS J18455+0530,  ADS~11640,  or FIN~332.   Other identifiers  are
HIP~92027,   HD~173495,   HR~7048.   The   outer   2.5$''$  pair   A,B
(STF~2375AB), discovered by W.~Struve in 1825, consists of two similar
A1V  stars, each  of  them  itself being  a  close binary.   W.~Finsen
discovered the subsystems in 1953 using an eyepiece interferometer and
called   them   ``Tweedledum  and   Tweedledee''   because  of   their
similarity. Rich and  at times controversial observational history
of the ``Tweedles'' is  related by \citet{Msn2010}. Ironically, SIMBAD
does not list this pivotal paper among references on this object.

\begin{table}
\center
\caption{Components of FIN 332}
\label{tab:par2}
\begin{tabular}{l c c cc  }
\hline
Parameter  & Aa    & Ab   & Ba   & Bb \\
\hline
$V$ (mag) & 6.98 & 7.38  & 7.47 & 7.47 \\
${\cal M}$ (${\cal M}_\odot$) & 2.46 & 2.20 & 2.14 & 2.14 \\
\hline
\end{tabular}
\end{table}

The Gaia parallaxes are 4.69$\pm$0.47  mas for A and 5.48$\pm$0.30 mas
for  B  \citep{Gaia}.  Both  are  inaccurate,  considering the  binary
nature of the  sources.  I adopt the dynamical parallax  of 6.0 mas in
the  following.   Individual  magnitudes in  Table~\ref{tab:par2}  are
derived from the $V$ magnitudes of  A and B measured by Gaia (6.41 and
6.72 mag) and the $\Delta y$  of the close pairs measured at SOAR, 0.4
and  0.0 mag  for Aa,Ab  and Ba,Bb,  respectively.  Gaia  measured the
effective temperature of A and B  as 9613 and 9169 K, corresponding to
spectral  types  A0V  and  A1V.   Masses  derived  from  the  absolute
magnitudes (assuming no extinction), from  2.14 \msun for Ba and Bb to
2.46 \msun for Aa, match masses expected for these spectral types. The
combined color of  all 4 stars $V  - K = 0.17$ mag  corresponds to the
spectral type A2, while thea ctual spectral types imply $V-K \sim 0.1$
mag.  Hence the interstellar extinction is indeed negligible.

\subsection{Orbits of FIN 332}
\label{sec:orb2}

\begin{figure*}
\plotone{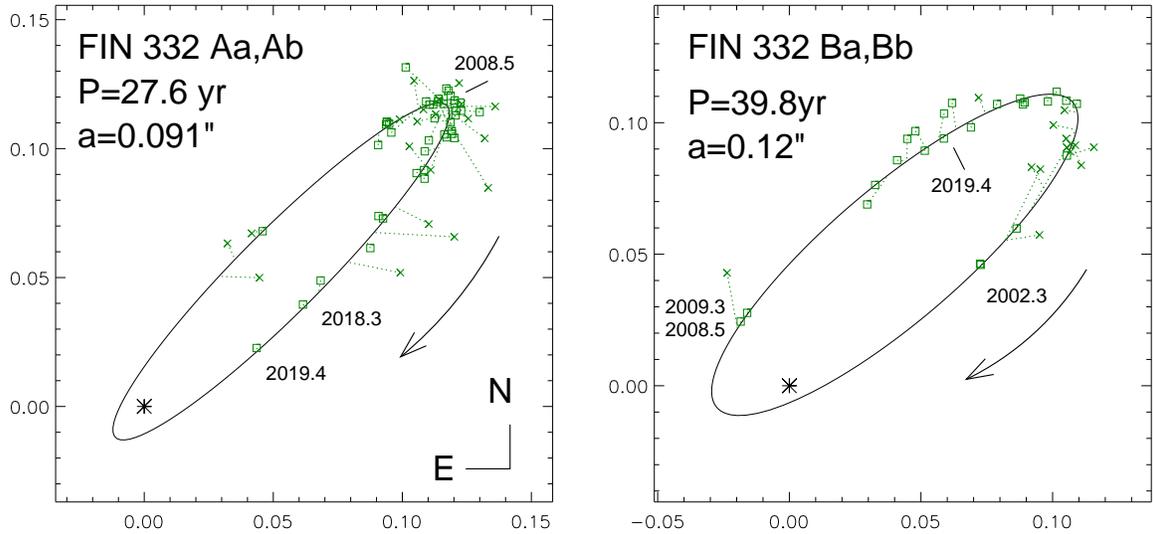}
\caption{Orbits  of  inner subsystems Aa,Ab and Ba,Bb in FIN 332. 
\label{fig:orb2} 
}
\end{figure*}

\begin{figure}
\plotone{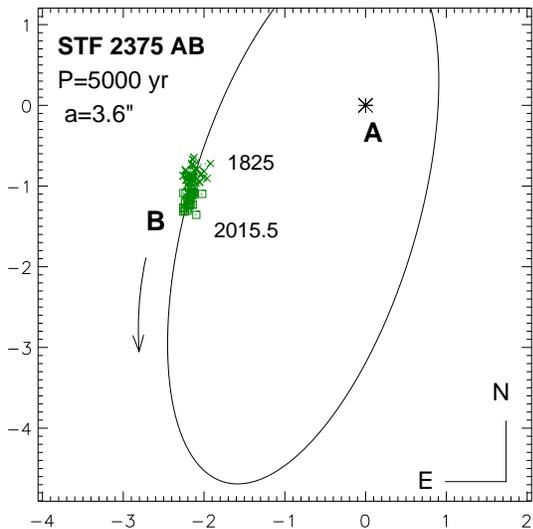}
\caption{Notional outer orbit of STF 2375 AB. Only a small arc is
  covered since the discovery of this pair in 1825. The orbital
  elements are given in Table~\ref{tab:orb}.
\label{fig:orb2a}
}
\end{figure}

\citet{Msn2010}  determined the  first reliable  orbits  of subsystems
Aa,Ab and  Ba,Bb (periods  of 27.02 and  38.6 yr,  respectively) after
critical evaluation  and correction of the existing  data.  They based
orbits  only on  speckle  interferometry and  ignored historic  visual
measurements,  as well  as  interferometric measurements  made by  the
author  at  the 1-m  telescope  \citep[e.g.][]{Tok1982}.  Since  then,
additional  measurements  at  the  6-m  telescope  were  published  by
\citet{Bag2013},   and    the   system   was    monitored   at   SOAR.
\citet{Msn2018c} updated the orbits to  periods of 27.74 and 39.92 yr.
The   orbits  listed   here  in   Table~\ref{tab:orb}  and   shown  in
Figure~\ref{fig:orb2} are  very similar to  those of \citet{Msn2018c}.
However, no errors  are quoted by Mason, hence  the need to re-compute
the orbits here.

In fitting the orbits, small  errors of 2\,mas (hence high weight) are
adopted for  speckle measurements  at SOAR and  at the  6-m telescope.
Other speckle data from 4-m  telescopes are assigned errors of 5\,mas,
the   measurements  from  smaller   telescopes  have   larger  errors.
Interferometric measurements  by W.~Finsen are used with  a low weight
(30\,mas error)  for better definition  of orbital periods,  while the
micrometer data are ignored.  The time base of 65.7 yr (from 1953.7 to
2019.4) covers 2.4 orbital periods  of Aa,Ab and 1.7 periods of Ba,Bb.
The  SOAR  measurement  of  Ba,Bb  in  2009.26  is  added  (originally
published   as  unresolved)   and  the   measurement  in   2008.55  is
reprocessed.  The weighted rms residuals are 3 mas for Aa,Ab and 2 mas
for Ba,Bb. Note  that, despite similar input data,  the semimajor axis
of Ba,Bb  is determined  with a 10$\times$  larger error,  compared to
Aa,Ab.  This  is because the  elements $T, e,  a, \omega, i$  of Ba,Bb
have  large mutual  correlations. Little  can be  done to  improve the
orbit because Ba,Bb is now far from the periastron.  Its periastron in
2005 has not been  covered, and the next one will be  in 2045.  On the
other hand, Aa,Ab is now closing down (next periastron in 2021.6), and
its regular monitoring will soon further constrain the orbit.

The  estimated masses  and  the  orbit of  Aa,Ab  yield the  dynamical
parallax  of   6.0\,mas,  while  the  less  certain   orbit  of  Ba,Bb
corresponds  to the dynamical  parallax of  6.4\,mas.  If  the element
$\omega$ of Ba,Bb is fixed at $310^\circ$ (only one standard deviation
off the best  fit), the semimajor axis of  Ba,Bb increases to $0.113''$
and the dynamical parallax becomes 6.0\,mas. Forcing $\omega$ does not
affect the period. The ratio of inner periods in this quadruple system
is 1.44$\pm$0.02.

The  outer  period estimated  from  the  separation  between A  and  B
($2.6''$ or 433 au)  and the mass sum is of the  order of 3\,kyr.  The
position angle of A,B has  increased from 108$^\circ$ at its discovery
in  1825 to 120$^\circ$  now.  Notably,  the outer  pair has  a direct
motion, while both inner orbits are retrograde. The short observed arc
does not constrain the outer orbit.  A notional orbit with a period of 5
kyr is provided as illustration in Figure~\ref{fig:orb2}.

\subsection{Architecture of FIN 332}
\label{sec:FIN332disc}

FIN~332 is a  typical 2+2 quadruple system of   $\epsilon$~Lyr type
\citep{Tok2008}.    All  four   stars  have   comparable   masses  and
luminosities,  meaning that they  were not  chosen randomly  from some
general  mass distribution.  Like  in many  other 2+2  quadruples, the
periods  of  the inner  subsystems  are  comparable. Most  remarkably,
however, the two inner orbits also have similar orientation in the sky
(Figure~\ref{fig:orb2}),  similarly   large  eccentricities,  and  similar
orientation of the lines of apsides. Mutual inclination between
orbits of Aa,Ab and Ba,Bb is either $16.1^\circ$ or $141.5^\circ$
(the orbital nodes are ambiguous). 

Relative inclinations of the inner subsystems to the tentative outer
orbit are either $56-60^\circ$ or $135-150^\circ$. These numbers are
only indicative, given the uncertain orbit of A,B. However, approximate
coplanarity between outer and inner orbits is excluded by the
apparent counter-rotation. Large mutual inclination leads to
Kozai-Lidov cycles which can drive inner eccentricity to large
values. Indeed, both inner orbits are very eccentric. 

Components of this  system have fast axial rotation  of $\sim$150 \kms
typical of  early-type A stars,  making determination  of spectroscopic
orbits unlikely. The  average RV is $-19.2 \pm  0.9$ \kms according to
\citet{Gontcharov2006}. The  PMs measured by Gaia can  be distorted by
orbital  motion in  the subsystems.  This is  less likely  for  B with
equal-brightness components. Its Gaia  PM, corrected for the motion in
the  A,B orbit  (B moves relative to A at  $(+0.8, -3.2)$  mas/yr) leads  to the
center-of-mass  PM of  $(16.4, 0.6)$  mas/yr.  The mean  PM of  component A
deduced  from the  Gaia  and Hipparcos  positions  is $(+16.2,  +1.6)$
mas/yr, corresponding to the center-of-mass PM of $(15.8, 0.0)$
mas/yr. The two estimates of the center-of-mass PM are close to each
other. Their average, RV, and dynamical parallax of 6 mas lead to the
Galactic velocity of $(U,V,W) = (-18.3, -6.3, -12.4)$ \kms. This
coresponds to  young disk population, but cannot be associated with
known kinematic groups.  

\section{Summary and Discussion}
\label{sec:disc}

Discovery by \citet{Zasche2019}  of potential resonances between inner
binaries  in  doubly  eclipsing  systems  motivated  this  work.   The
accuracy  of inner  periods in  the  two hierarchies  studied here  is
sufficient to  prove that the  period ratios are  measurably different
from  rational  numbers:  2.87$\pm$0.07 and  1.44$\pm$0.02.   However,
period ratios  in resonant  multi-planet systems and  doubly eclipsing
binaries  also  differ  from  exact  rational numbers  by  1-2\%,  and
differences    of    this    order    are   expected    from    theory
\citep{Tremaine2020}. Note that  Tremaine's analysis of close binaries
on circuar orbits is not  directly applicable to our quadruple systems
with eccentric inner orbits.

Notional  outer orbits allow  us to  estimate the  ratio of  inner and
outer  semimajor   axes  that   governs  the  strength   of  dynamical
interaction between inner and outer orbits, $\epsilon = {\rm max}(a_1,
a_2)/a_3$.  This  parameter is  about 0.08 and  0.03 for  GJ~225.1 and
FIN~332, respectively. Therefore,  interaction between inner and outer
orbits  is far  from  being negligible.  Dynamical  analysis of  these
systems is beyond the scope  of this paper which focuses on assembling
the observationa data -- orbits and masses.

It is  instructive to speculate  on the formation mechanisms  of these
hierarchies.  Their structure is  far from  being chaotic:  the orbits
show  some mutual  alignment, and  the  masses of  components in  each
system are  comparable (except  GJ~225.1~E). It appears  unlikely that
these  hierarchies  experienced  strong  internal or  external  (in  a
cluster)  dynamical interactions.  Multiple systems  surviving chaotic
dynamics  are different:  they have  eccentric and  misaligned orbits,
while their masses are not so well correlated \citep{ST02}.

Similar masses imply that components of these hierarchies accreted gas
from a  common source. Most  likely, these systems formed  in relative
isolation  by collapse  of an  overdensity (core  or  filament).  This
scenario  was  proposed as  formation  mechanism  of wide  hierarchies
composed of similar solar-type stars \citep{Tok2020}; their wide outer
separations  imply   absence  of  close   neighboring  stars.   Recent
hydrodynamical simulations of  isolated cloud collapse show successive
formation  of protostars,  migration  of binaries  to shorter  periods
driven by accretion, and formation  of outer companions that, in turn,
accrete and migrate inward \citep{Lee2019,Kuffmeier2019}.

Accretion-driven  migration  is  a  viable  mechanism  to  form  close
(spectroscopic)  binaries   \citep{TM20}.  In  this   respect,  it  is
important  to  note that  some  2+2  hierarchies  contain inner  close
subsystems.  For  example,  the  wide  2+2  hierarchy  ADS  9716  (HIP
76563/76566,  outer  projected  separation  1.6  kau)  contains  inner
spectroscopic subsystems with periods of  3.3 and 14.3 days and counts
six stars  in total \citep{Tok1998}. Presence of  close inner binaries
is a strong (albeit  indirect) argument for accretion-driven evolution
of stellar hierarchical systems.

Successive  formation of  companions  during collapse  of an  isolated
cloud and  their inward migration matches the  architecture of compact
planetary-like hierachical systems  where all orbits are approximately
coplanar, their eccentricities are moderate, and the ratios of periods
are not  extreme, e.g.   the 3+1 quadruple  HD~91962 \citep{Tok2015a}.
This    architecture    is    typical   for    low-mass    hierarchies
\citep{twins}. It matches the properties of GJ 225.1, except that this
is a 2+2 hierarchy.  The low-mass  companion E could be formed by disk
fragmentation; a massive and unstable  disk around C could be produced
by a  late accretion  burst, when  most of the  mass has  already been
accreted by the first three stars  A, B, and C. This scenario explains
the  low  mass ratio  of  C,E. It  could  work  for other  hierarchies
containing inner  subsystems with  low mass ratios,  e.g. $\alpha$~Gem
(HIP  36850),  a visual  pair  where  each  star is  a  single-lined
spectroscopic binary.

The architecture  of  the  more  massive  quadruple  system  FIN~332  (the
Tweedles)  is different:  its inner  and outer  orbits  are definitely
misaligned,  although  the  inner   orbits  might  still  be  mutually
aligned. Statistically, there is  no alignment between inner orbits of
resolved 2+2 quadruple systems, as can be inferred from the comparable
numbers of  apparently co- and counter-rotating inner  pairs.  In this
sense,   FIN~332   is  atypical.    Similarity   of  orientation   and
eccentricities      of     its      inner     orbits      is     truly
remarkable. Hypothetically,  such quadruples resembling $\epsilon$~Lyr
could  form by  the  outside-in hierarchical  collapse,  possibly
triggered by collision  \citep{Whitworth2001}. However, the similarity
of component's masses in such quadruples still suggests accretion from
a common source. Accretion helps  to shrink the initially wide (on the
order of Jeans length?) stellar hierarchies to their actual
size. Overall, larger and more massive hierarchies are less aligned in
comparison to their smaller and less massive counterparts \citep{moments}. 

The two hierarchical systems presented here could be studied in detail
owing to the happy  coincidence of their parameters (separations, mass
ratios, distance) with past and current observational capabilities and
the  existence of  adequate  time coverage.   Continued monitoring  of
other hierarchical  systems and data  from large surveys  will provide
material for study of their dynamics and origin.

The work  of Tokovinin  is supported by  NOIRlab, which is  managed by
Association of  Universities for Research in Astronomy  (AURA) under a
cooperative agreement with the National Science Foundation.








\end{document}